\documentclass[aps,prd,twocolumn,groupedaddress,showpacs]{revtex4}

\usepackage[dvips]{graphicx}
\usepackage{amsmath, amssymb}

\newcommand{\mathsym}[1]{{}}

\begin{document}

\title{Harmonic Gravitational Wave Spectra of Cosmic String Loops in the Galaxy}

\author{Matthew R. DePies}
\email{depies@phys.washington.edu}
\affiliation{University of Washington, Department of Physics, Seattle WA, 98195}

\author{Craig J. Hogan}
\email{craighogan@uchicago.edu}
\affiliation{University of Chicago and Fermilab, PO Box 500, Batavia, IL 60510}

\date{\today}

\begin{abstract}
A new candidate source of gravitational radiation is described: the nearly-perfect harmonic series from individual loops of cosmic string.  It is argued that theories with light cosmic strings give rise to a population of  numerous  long-lived stable loops, many of which cluster gravitationally in galaxy halos along with the dark matter.  Each cosmic string loop produces a spectrum of discrete frequencies in a nearly perfect harmonic series, a fundamental mode and its integer multiples. The gravitational wave signal from cosmic string loops in our Galactic halo is analyzed numerically and it is found that the for light strings, the nearest loops typically produce strong signals which stand out above confusion noise from Galactic binaries.  The total population of cosmic string loops in the Milky Way also produces a broad signal that acts as a confusion noise.   Both signals are enhanced by the clustering of loops gravitationally bound to the Galaxy, which significantly decreases the average distance from the solar system to the nearest loop.  Numerical estimates indicate that for dimensionless string tension $G\mu/c^2 < 10^{-11}$,  many loops are likely to be found in the Galactic halo.  Lighter strings, down  to $G\mu/c^2\approx 10^{-19}$, are detectable by the \textit{Laser Interferometer Space Antenna} (LISA).   For these light strings, the fundamental and low-order harmonics of typical loops often lie in the band where LISA is sensitive, 0.1 to 100 mHz.  The harmonic nature of the cosmic string loop modes leaves a distinct spectral signature different from any other known source of gravitational waves.
  \end{abstract}


\maketitle

       \section{Introduction}
       
Topological defects produced during cosmic phase transitions are a standard component of many field and brane theories and cosmologies.  Often they take the form of macroscopically extended  classical one dimensional objects, with microscopic radius,  called cosmic strings.  The cosmological evolution of strings forms a population of quasi-stable loops that lose energy mainly by gravitational waves~\cite{hin,quashnock,an,ho87,ho84,garfinkle,silk,hogan84,ki,sak06,vi81,batt,turok,allen01,vashvil85,linde,vil,ba,va84,vil85,vi05,pol,dv,sas,jones1,pol06,tye06,fir05,jones,dubath07,all94,allen94,all95,casper95,hindstuck08,turok1984}.   This paper describes a new way in which these gravitational waves may be observed: a regime in which radiation from individual loops can appear as perfect harmonic series in an observed frequency spectrum.

String loops  oscillate, radiate gravitational waves and shrink until they completely decay.  The  center of mass speed of a loop when it forms is of the order of unity. After it stops interconnecting with the rest of the string network, a loop's velocity decays inversely proportional to the scale factor.  At late times, the primordial velocity is negligible and the loop population clusters in almost the same way as the dominant cold, collisionless dark matter~\cite{note}.  In the case of light strings, for which loops are small and numerous, a galaxy halo can contain a very large number of loops~\cite{chernoff07}.  They are concentrated to high density, so the mean distance to the nearest loop is much smaller than the cosmic mean.

A cosmic string loop  produces a spectrum of discrete frequencies~\cite{vil,anderson} which may be detectable if it is close enough. The spectrum of any loop is given by a sum over a nearly perfect harmonic series of frequencies $f_n=2nc/L$, where $L$ is the length of the string loop.  This distinctive property is unlike any other astrophysical source of gravitational waves and if observed, would provide convincing evidence of the existence of cosmic strings, as well as detailed information about their astrophysical behavior.

The power from each discrete mode $n$ of a loop is given by,
\begin{equation}
\dot{E}_n=P_n G \mu^2 c,
\end{equation}
where $\mu$ is the mass (energy) density of the string.
We define $\gamma=\sum{P_n}$.  Numerical simulations indicate $\gamma$ is approximately 50 to 100.  We use a value of 50 for this study.
The power in each mode depends upon the particular oscillation pattern of a string loop, but the general solution of a sum of harmonic modes does not depend upon any particular model.     For the illustrative estimates in this paper we assume a very simple model, again motivated by numerical estimates: the mean power of an ensemble scales as $\langle P_n \rangle \propto n^{-4/3}$.    
Loops clustered around the Milky Way halo, most of which are too distant to detect individually, taken together create an unresolved background of gravitational waves, akin to the white dwarf binaries~\cite{nel,farm}. We also estimate this confusion background.    

The new effects are most important for cosmic string loops with lower string tensions $G\mu/c^2<10^{-12}$~\cite{depies,ho} and for stable loops formed at a significant fraction, $\alpha=0.1$, of the horizon~\cite{depies,ho,vanch08}.  In this situation, we find that cosmic strings are an important new source  at   the frequencies 0.1 Hz to 10 mHz~ band where \textit{Laser Interferometer Space Antenna} (LISA)~ \cite{lisa,shane}  will be most sensitive.  The extension into this regime is all the more relevant given the current limits set on cosmic string tensions via SDSS, WMAP, and millisecond pulars~\cite{jenet,wy}.  Since the results depend on the size distribution of stable loops~\cite{vanch08,hindstuck08}, a number of possibilities are considered in the parameter studies here.

  \section{Distribution of Loops}
     
The average number density of loops between size $L$ and $L+dL$ is $n(L)dL$, computed numerically.  The loops are assumed to be clustered around galaxies in the same way as the dark matter.   The number density in the galaxy is matched to the dark matter halo, given by the NFW density distribution $\rho_{NFW}(r)$~\cite{NFW95,klypin02,binney, chandra, allen}.    The number density is then a function of the length of loops and distance from the center of the galaxy $n(L,r)$.

\subsection{Loop Density}

We start by figuring out the size of the loops at the fundamental frequency:
\begin{equation}
L=2c/f_1
\end{equation}
Using the one-scale model~\cite{ringeval07,rocha08}, loops form with size $L \approx \alpha c H^{-1}(t)$ and start to decay at a rate $\dot{L}=-\gamma G \mu$.  Loops created at a time $t_c$ will be of the size $L(t_c,t)$ at the time $t$. 
At the present time $t_o$ we find the fundamental frequency given by:
\begin{eqnarray}
f_1 &=& \frac{2c}{\alpha c H^{-1}(t_c)- \gamma G \mu (t_o-t_c)/c}.
\end{eqnarray}
The number density at $t_o$, for time of creation $t_c$  is given by:
\begin{equation}
n(t_o, t_c)=\frac{ N_t }{\alpha} \left( \frac{H(t_c)}{c}\right)^3 \left( \frac{a(t_c)}{a(t_o)}\right)^3,
\end{equation}
Using the time $t_c$ we solve numerically to find the number density.

To illustrate with typical numerical values: the average distance between galaxies is 5 Mpc, the number of loops within the galaxy for $G\mu=10^{-12}$ and a fundamental frequency of 1 mHz is approximately $10^4$.

        \subsubsection{Density in the Galaxy}
        
The density of galactic string loops is matched to the dark matter distribution in the Milky Way using the NFW density $\rho_{NFW}(r)$:
\begin{eqnarray}
\rho_{NFW}(r) &=& \frac{\rho_s}{x(1+x)^2},\\
x &=& \frac{r}{r_s},
\end{eqnarray}
where $\rho_s$ and $r_s$ are determined by observation.  Representative values are given by~\cite{klypin02} using the favored model, e.g. $r_s$=21.5 kpc.  From above we see that for $r<<r_s$, $\rho \propto r^{-1}$ and for $r>>r_s$, $\rho \propto r^{-3}$.  
      
From the previous example using strings of tension $G \mu=10^{-12}$, $N=10^4$, $r_t/r_s = 10$, and $r_s=21.5$ kpc, we find at the earth's distance from the galactic center, $r=8$ kpc, a value of $n \approx 9.4 \times 10^{-2}$ kpc$^{-3}$, or $n^{-1/3} \approx 2.2$ kpc, an ``average'' distance to the nearest loop. The nearst loop is likely closer as the distribution is not constant over this range of position.

 The number density of loops in the Milky Way is strongly influenced by the size and mass density of the cosmic string loops.  Tables~\ref{table1}--\ref{table3} give results for the number of loops in the Milky Way for various parameterizations of the string loops.  The results clearly show the large predominance of loops for very light and large strings.
    
    \begin{table}[H] 
 \caption{\label{table1}Number of Loops in the Milky Way for the given fundamental frequencies and $\alpha$=0.1}
 \begin{ruledtabular}
 \begin{tabular}{c c c c c}
  
          &  &          &     \\
  $G\mu$   &   $f$(Hz) $ 10^{-4}  $          &       $ 10^{-3}  $     &    $ 10^{-2}$     &     $ 10^{-1}  $                    \\   
\hline
 $10^{-11}$ \vline &     3 $\times 10^2$&         3 $\times 10^2$          &      3 $\times 10^2$        &       3 $\times 10^2$ \\

$10^{-12}$  \vline &  1$\times10^4$              &          1$\times10^4$      &     1$\times10^4$       &       1$\times10^4$                   \\
 
$10^{-13}$  \vline & 3 $\times10^{5}$& 3 $\times10^{5}$  & 3 $\times10^{5}$   &3 $\times10^{5}$       \\
 
$10^{-14}$  \vline & 9 $\times10^{6}$&  1$\times10^{7}$  &  1$\times10^{7}$   &  1$\times10^{7}$       \\

$10^{-15}$  \vline & 1$\times10^{8}$& 3 $\times10^{8}$  & 3 $\times10^{8}$   &3 $\times10^{8}$        \\

$10^{-16}$  \vline & 3 $\times10^{8}$& 4 $\times10^{9}$  & 9 $\times10^{9}$   & 1$\times10^{10}$        \\

 \end{tabular}
 \end{ruledtabular}
 \end{table}

  \begin{table}[H] 
 \caption{\label{table2}Number of Loops in the Milky Way for the given fundamental frequencies, varying $\alpha$ and G$\mu=10^{-12}$}
 \begin{ruledtabular}
 \begin{tabular}{c c c c c}
  
         &  &        &     \\
  $\alpha$        &   $f$(Hz) $ 10^{-4}  $     &       $ 10^{-3}  $   &    $ 10^{-2}$       &     $ 10^{-1}  $               \\   
\hline
 $10^{-1}$ \vline & 1$\times10^4$       &     1$\times10^4$        &     1$\times10^4$     &        1$\times10^4$              \\

$10^{-2}$  \vline & 3$\times10^3$        &         3$\times10^3$    &     3$\times10^3$     &        3$\times10^3$      \\
 
$10^{-3}$  \vline & 1$\times10^3$         &       1$\times10^3$    &    1$\times10^3$     &    1$\times10^3$           \\
 
$10^{-4}$  \vline & 4$\times10^2$        &          4$\times10^2$   &     4$\times10^2$  &       4$\times10^2$      \\

$10^{-5}$  \vline &  2$\times10^2$      &          2$\times10^2$    &     2$\times10^2$       &      2$\times10^2$                  \\
 
$10^{-6}$  \vline &  1$\times10^2$          &     1$\times10^2$      &      1$\times10^2$        &    1$\times10^2$           \\
  
 \end{tabular}
 \end{ruledtabular}
 \end{table}

  \begin{table}
 \caption{\label{table3}Number of Loops in the Milky Way for the given fundamental frequencies, varying $\alpha$ and G$\mu=10^{-16}$}
 \begin{ruledtabular}
\begin{tabular}{c c c c c}
  
          &  &         &     \\
  $\alpha$        &   $f$(Hz) $ 10^{-4}  $      &       $ 10^{-3}  $   &    $ 10^{-2}$       &     $ 10^{-1}  $                 \\   
\hline
 $10^{-1}$ \vline & 3$\times10^{8}$& 4 $\times10^{9}$  & 9$\times10^{9}$   &1$\times10^{10}$      \\

$10^{-2}$  \vline & 1$\times10^{8}$& 1$\times10^{9}$  & 3$\times10^{9}$   &4$\times10^{9}$       \\
 
$10^{-3}$  \vline & 3 $\times10^{7}$& 4 $\times10^{8}$  & 9 $\times10^{8}$   &1$\times10^{9}$      \\
 
$10^{-4}$  \vline & 1 $\times10^{7}$& 1$\times10^{8}$  & 3$\times10^{8}$   &3$\times10^{8}$     \\

$10^{-5}$  \vline & 3 $\times10^{6}$& 4$\times10^{7}$  & 9$\times10^{7}$   &1 $\times10^{8}$      \\
 
$10^{-6}$  \vline & 1 $\times10^{6}$& 1$\times10^{7}$  & 3$\times10^{7}$   &3$\times10^{7}$      \\
  
 \end{tabular}
 \end{ruledtabular}
 \end{table}

  \section{Power from each loop}

A loop radiates power in each mode $n$ modeled by,
\begin{equation}
\dot{E}_n(r') \propto \frac{\dot{E}_n}{r'^{2}},
\end{equation}
where $\dot{E}_n$ is power radiated per mode from a loop.  The total power is 
\begin{equation}
\dot{E}=\gamma G \mu^2=\sum P_n G \mu^2,
\end{equation}
where $\gamma$ is found to be between 50 and 100, and the $P_n$ are power coefficients of the individual modes.  Note that the power radiated is independent of the loop size.  To first order we ignore directionality of the emitted gravitational radiation.

Each loop has a spectrum given by a set of power coefficients, $P_n$, and the average sum for the population $\langle\sum{P_n}\rangle\approx \gamma$ is fixed by the statistical properties of the strings.  This is an average over a wide array of loops:  individual loops vary from this.  For most loops the most power is in the fundamental mode and the lowest modes, while the higher modes have significantly reduced power.

For a given mode the quality factor is given by,
\begin{equation}
Q_n=\frac{4 \pi n}{P_n G \mu}.
\end{equation}
For light strings with $G\mu=10^{-12}$ we find,
\begin{equation}
Q_n \approx 10^{12} \frac{n}{P_n}.
\end{equation}
For the fundamental mode $P_n$ is of order unity at most, so $Q$ is large, and increases with higher $n$.  Thus the harmonic series is almost perfect and the lines extremely narrow. This justifies the use of delta functions in other applications.

Given the extremely large value of $Q_n$  the signal from each mode in the loop is practically constant for the duration of observation with respect to decay.  Potential frequency shifts due to path differences from gravitational lensing, gravitational doppler effect, Newtonian acceleration in the Galaxy, and relative motion have not been analyzed~\cite{dubath07}.


  \section{Strain Produced by Gravitational Radiation}

The relative weakness of gravitational waves from cosmic strings allows the use of linearized gravity.  
Assuming plane wave solutions, we find for the energy flux:
\begin{equation}
F_{gw}=\frac{\pi}{4 G} f_n^2 h_n^2,
\end{equation}
where $h_n$ is the root-mean-square strain, and using $c=1$.
Inserting $c$ we find :
\begin{eqnarray}
h_n &=& \sqrt{P_n} \frac{c}{\pi} \frac{G \mu}{c^2} \frac{1}{r f_n},\\
h_n &=& 3.095 \times 10^{-12} \sqrt{P_n} \; [G \mu (c=1)] \times \nonumber \\
     & & \;    \left(\frac{1 \text{Hz}}{f_n}\right)  \left( \frac{1 \text{kpc}}{r} \right).
\end{eqnarray} 
For $G \mu=10^{-12}$, $f$=1 mHz, and $r$=1 kpc, we find $h_1 \approx 10^{-21}$, within the detection limits of LISA.

   
   \section{Spectrum of a Single Loop}
   
Using previous results for the strain produced by a gravitational wave, we can add the harmonic time dependence to find the strain at each mode for the result:
\begin{equation}
h_n(t)=\frac{c\sqrt{P_n}}{\pi f_n} \frac{G \mu}{c^2} \frac{1}{r}\,e^{-i 2 \pi f_n t}
\end{equation}
Again, the $P_n$ are the ``power coefficients''  estimated to scale in the mean as $n^{-4/3}$.  Recall that the sum of coefficients is given by $\sum P_n=\gamma$.  In practice most of the power is in the fundamental mode, so the the power from lower modes drops off, allowing truncation of the sum at a reasonable value of $n$.   As an example, we set teh amplitudes equal to the sample mean:
\begin{equation}
P_n= P_1 n^{-4/3},
\end{equation}
and $P_1=18$.  At $n=20$ we find $\sum P_n \approx 45$, giving   a reasonable estimate.
For a single loop the total strain will be a sum of all the modes,
\begin{equation}
h(t)=\sum_{n=1}^{\infty} h_n(t).
\end{equation}
In practice a suitable cutoff is used to facilitate numerical computations.

            \subsection{Calculation of Single Loop Spectrum using Discrete Fourier Transform}

Since the signal detected by LISA will be as a discrete set of points, it is useful to take a discrete Fourier transform of a sample signal.  In general LISA has a sampling rate of $f_{samp}=10$Hz, which leads to large numbers of data points for integration times of one to three years.  This large number of data points also leads to a very clear signal due to the harmonic nature of the gravitaional waves produced by the cosmic strings.

Given the signal $h(j)$ where $j$ is an integer and the $j^{th}$ data point corresponding to $h(t)=h(j \Delta T)$, we define the discrete-time Fourier transform as~\cite{teul},
\begin{equation}
\label{eq:dft}
\tilde{h}(k)=\frac{1}{\sqrt{T}} \sum_{j=1}^N h(j)e^{-i 2 \pi \frac{(j-1)(k-1)}{N}}.
\end{equation} 
$N$ is the total number of data points and $k$ is the integer value in Fourier frequency space.  Note the introduction of the square root on the integration time $T$, which is standard in the sensitivity curves for LISA~\cite{shane1}.

The gravitational wave signal from the cosmic strings has the form,
\begin{eqnarray}
h(j)=\sum_{n=1}^{\infty} \frac{c\sqrt{P_n}}{\pi f_n} \frac{G \mu}{c^2} \frac{1}{r} \sin(2 \pi j \omega_n/N),
\end{eqnarray}
where $\omega_n$ is not the physical frequency, but the discrete time frequency, which is different than the physical frequency given by $f_n=2nc/L$, and $P_n$ is the magnitude of the $n^{th}$ power mode.  Using $P_n=P_1 n^{-4/3}$ this is then inserted into Eq.~\ref{eq:dft}.

\begin{figure}
   
	 \includegraphics[width=.46\textwidth]{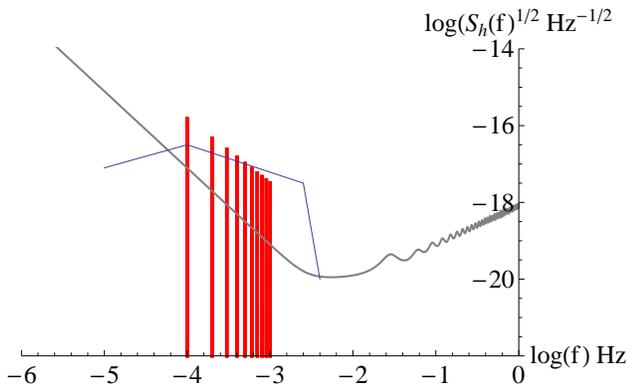}
  \caption{\label{fig51}Plot of the cosmic string loop spectrum for $G\mu=10^{-11}$, $\alpha$=0.1, at a distance of $r=7$ kpc.  Observation time is $T=1$ year, the sampling rate is 0.1 Hz, $P_1=18$, and the fundamental frequemcy is $f_1=10^{-4}$ Hz.  The first 10 modes are shown.  This signal stands above the confusion noise from galactic binaries which is shown on the graph.}
		\end{figure}

Investigations of the parameter space of the loops indicates the heaviest individually detectable loops are of tension $G\mu \approx 10^{-10}$ for $\alpha=0.1$, due to their reduced numbers in the Milky Way, see Table~\ref{table1}.
 Shown in Fig.~\ref{fig51} is the largest likely signal from an individual loop at the frequency $10^{-4}$ Hz.  This signal stands above the noise from binaries. 

For the large loops, $\alpha=0.1$, it is found that the lightest individually detectable loops  have tension $G\mu>10^{-16}$.

\begin{figure}
	 \includegraphics[width=.48\textwidth]{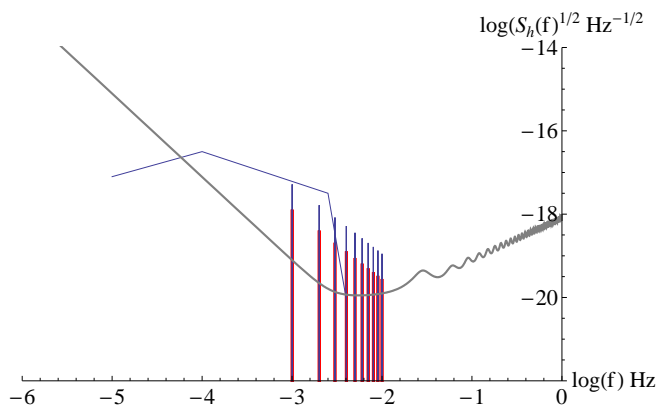}
  \caption{\label{fig52}Plot of two cosmic string loop spectra with $G\mu=10^{-12}$: the ``thin'' spectrum $\alpha$=0.1 at a distance $r=2.2$ kpc from the solar system, and the ``thick'' spectrum $\alpha=10^{-5}$ with $r$=8.9 kpc.}
		\end{figure}

The average size $\alpha$ of the loops   as a fraction of the horizon when they formed   affects the number of loops currently radiating.  For large $\alpha$ the number of loops remaining is large, thus at a given frequency the distance to the nearest loop is shorter.  This is indicated in Fig.~\ref{fig52}.  Note the amplitude varies as $r^{-1}$, which shows only a small difference on our log plots.  
The total number of loops is significantly larger for large $\alpha$.

\begin{figure}
	 \includegraphics[width=.48\textwidth]{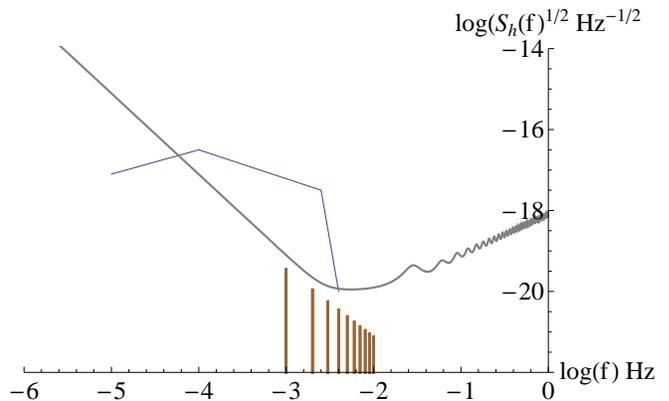}
  \caption{\label{fig53}Plot of the Fourier spectrum of cosmic string loops with tension $G\mu=10^{-16}$ and a distance of $r=0.066$ kpc.  The fundamental frequemcy is $f_1=1$ mHz and the first 10 modes are shown.  Note the heavier loops ($G\mu=10^{-12}$) have a much larger signal, in spite of their greater distance from the solar system.  The confusion noise from galactic binaries is also shown.  At this fundamental frequency, the loop is not detectable, but shifted to higher frequencies it is.}
		\end{figure}

The increase in the radiated power of the heavier string loops results in greatest variation in signals potentially detectable.  Fig.~\ref{fig53} shows the  effect for string of tension $G\mu=10^{-16}$.   The heavier string loops are, on average, farther away but their larger output makes up for increased distance.

The signal to noise for a loops is given by the standard formula for a periodic source~\cite{maggtext},
\begin{equation}
SNR^2=|h_o|^2 \frac{T}{S_n(f)},
\end{equation}
where $T$ is the integration time and $S_n(f)$ is the noise spectral density of the detector.  For LISA the curve for $S_n(f)$ is given in~\cite{shane1}.

As an example we find for $G\mu=10^{-12}$, $\alpha=0.1$, and $f_1=10^{-3}$ that the $SNR\approx 10^5$.


   \section{Total Galactic Signal}

Another key ingredient is the gravitational wave signal from loops within the galactic halo.  These loops are not resolvable on an individual basis, but do contribute a significant gravitational wave signal.  The isotropic and stationary background from both current and evaporated loops has been calculated previously.  To this background we add the signal from loops within the dark matter halo of the milky way, which is not isotropic from the solar system.  

  We sum over distances in bins of $\Delta r'$ from the earth out to the edge of the galaxy.  The total flux $F^{net}$ of gravitational wave energy at a given frequency from strings of frequency $f_k=2k/L$ in the $k^{th}$ mode, within volume $dV'$ a distance $r'$ from the Earth is given by,
  \begin{eqnarray}
F_{gw}(f_k)^{net}= \int F_k^{gw}(r')\; n(\textbf{x}',L)\; d^3\textbf{x}'. 
\end{eqnarray}
Here we use $k$ to label the modes to avoid confusion with the number density $n(L)$.

From the total flux at a given frequency we find the strain,
 \begin{eqnarray}
h(f_n)= \frac{2}{f_n}\sqrt{\frac{G F_{gw}(f_n)^{net}}{\pi c^3}},
\end{eqnarray}  
which should be interpreted as the rms strain at a frequency $f_n$.

\begin{figure}
   
	 \includegraphics[width=.47\textwidth]{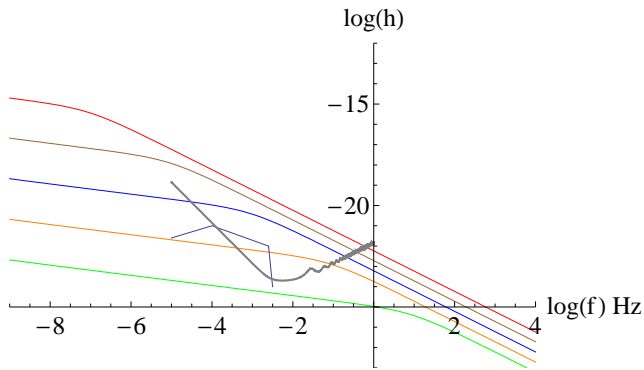}
  \caption{\label{fig63} Plot of the cosmic string loop strain spectrum for large loops $\alpha=0.1$ in the galaxy.  The top curve is of string tension $G\mu=10^{-12}$ and the bottom curve is $G\mu=10^{-20}$, in increments of $10^2$.  Also included are the LISA sensitivity curve with an integration time of 1 year, and the galactic white dwarf noise.  For each loop only the fundamental mode is included.}
		\end{figure}

\begin{figure}
   
	 \includegraphics[width=.47\textwidth]{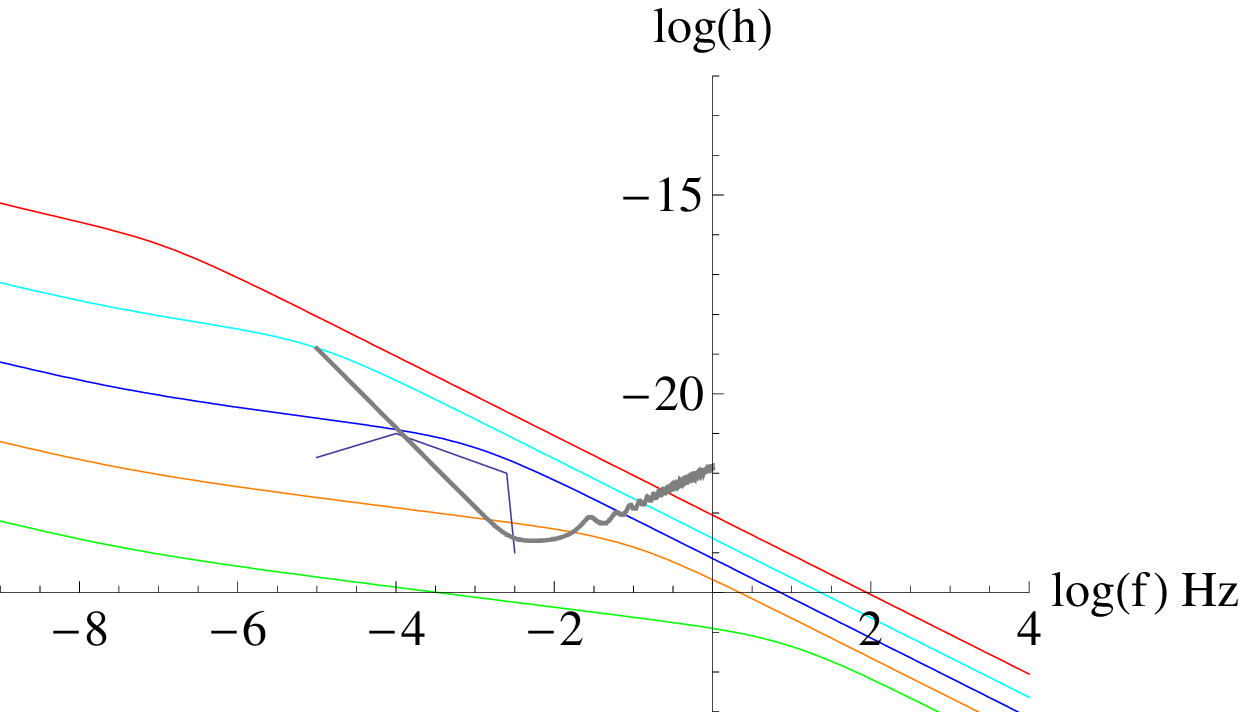}
  \caption{\label{fig64} Plot of the cosmic string loop strain spectrum for small loops $\alpha=10^{-5}$ in the galaxy.  The top curve is of string tension $G\mu=10^{-12}$ and the bottom curve is $G\mu=10^{-20}$, in increments of $10^2$.  Also included are the LISA sensitivity curve with an integration time of 1 year, and the galactic white dwarf noise.  For each loop only the fundamental mode is included.}
		\end{figure}

\begin{figure}
   
	 \includegraphics[width=.47\textwidth]{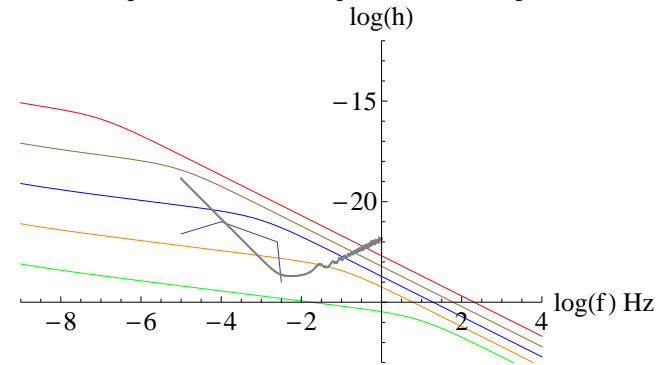}
  \caption{\label{fig65} Plot of the cosmic string loop strain spectrum for small and large loops in the galaxy.  It is assumed 90\% of the loops shatter into small loops of $\alpha=10^{-5}$.  The top curve is of string tension $G\mu=10^{-12}$ and the bottom curve is $G\mu=10^{-20}$, in increments of $10^2.$  Also included are the LISA sensitivity curve with an integration time of 1 year, and the galactic white dwarf noise.  For each loop only the fundamental mode is included.}
		\end{figure}

Results of numerical calculations are plotted in Figs.~\ref{fig63},~\ref{fig64}, and \ref{fig65}.

 \section{Conclusion}  

These results suggest a new way to observe light cosmic strings. They indicate that individual cosmic string loops are detectable due to the local concentration of   loops with   Galactic dark matter.  They  display a unique and distinctive harmonic spectrum which requires no special template fitting to detect, only a Fourier transform of the signal.  The large quality factor (of order 1/$G\mu$) ensures the frequency spectrum is  almost static over the several years of observation, further increasing detectability of the signal.

This study has taken a somewhat simplified approach to the distribution of the loops, simply matching them to the dark matter halo of the Milky Way.    The model of the loops is also oversimplified but adequate for the purpose of estimating detectability.

For the total galactic signal, we estimate  that Galactic string backgrounds are detectable by LISA down to $G\mu=10^{-19}$.  This is significantly more sensitive than previous results from the exragalactic stochastic background. Even though the broader cosmological impact of the loops is minimal, they still may provide a definite gravitational wave signal of new physics.

\begin{acknowledgments}
This research was funded by NASA grant NNX08AH33G at the University of Washington.
\end{acknowledgments}
   


\begin{thebibliography}{999}
\bibitem{note} Loops experience a gravitational wave rocket effect that tends to accelerate them over time. However, the typical rocket acceleration for the relevant loops is typically much smaller than  gravitational acceleration from dark matter halos, so this leads to only small differences between the dark matter and loop distributions.
\end{thebibliography}
\end{document}